\documentclass[pdflatex]{sn-jnl}

\jyear{2021}%

\theoremstyle{thmstyleone}%
\usepackage{lipsum}
\usepackage{mathtools}
\usepackage{cuted}
\usepackage[T1]{fontenc}
\usepackage[utf8]{inputenc}
\newcommand\sun{\odot}%

\usepackage{here}
\usepackage{scalerel}
\usepackage{float}
\usepackage{enumitem}
\usepackage{makecell}
\theoremstyle{thmstyletwo}%

\theoremstyle{thmstylethree}%
\begin{document}

\title{Relativistic stellar modeling with perfect fluid core and anisotropic envelope fluid}

\author*[1]{\fnm{A. C.} \sur{Khunt}}\email{ankitkhunt@spuvvn.edu}

\author[2]{\fnm{ V. O. } \sur{Thomas}}\email{votmsu@gmail.com}

\author[1,3]{\fnm{P. C.} \sur{Vinodkumar}}\email{p.c.vinodkumar@gmail.com}

\affil*[1]{\orgdiv{Department of Physics}, \orgname{Sardar Patel University}, \orgaddress{\city{Vallabh Vidynagar}, \postcode{388 120}, \state{Gujarat}, \country{India}}}

\affil[2]{\orgdiv{Department of Mathematics, Faculty of Science}, \orgname{The Maharaja Sayajirao University of Baroda}, \orgaddress{ \city{Vadodara}, \postcode{390 002}, \state{Gujarat}, \country{India}}}

\affil[3]{\orgdiv{Department of Physical Sciences}, \orgname{P. D. Patel Institute of Applied Sciences,
Charotar University of Science and Technology}, \orgaddress{\city{ Changa }, \postcode{388 421}, \state{Gujarat}, \country{India}}}


\abstract{We investigate the effect of density perturbations and local anisotropy on the stability of stellar  matter structures in general relativity using the concept of cracking. Adopting a core-envelope model of a super-dense star, we examine the properties and stability conditions by introducing anisotropic pressure to the envelope region. Furthermore, we propose self-bound compact stars with an anisotropic envelope as a potential progenitor for starquakes. We show how the difference between sound propagation in radial and tangential directions would be used to identify potentially stable regions within a configuration. Due to an increase in the anisotropic parameter, strain energy accumulates in the envelope region and becomes a potential candidate for building-up  quake like situation.  This stress-energy stored in the envelope region that would be released during a starquake of a self-bound compact star is computed as a function of the magnitude of anisotropy at the core-envelope boundary. Numerical studies for spherically asymmetric compact stars indicate that the stress energy can be as high as  $10^{50}$ erg if the tangential pressure is slightly more significant than the radial pressure. It is happened to be of the same order as the energy associated with giant $\gamma$-ray bursts. Thus, the present study will be useful for the correlation studies between starquakes and GRBs.}




\maketitle
\section{Introduction}\label{sec1}

Properties of  compact stars are one  the most sought-after topics in the field of astroparticle physics in recent times \cite{itoh1970hydrostatic, collins1975superdense, de2011self,horvath2021modeling}. Stability studies, starquakes, wobbling of stars and the processes such as giant gamma-ray burst etc are few examples. They release thousands  times more energy than supernovae. Their distinctive $\gamma$-ray emission lasts from fraction of seconds to  few minutes.  A typical burst releases as much energy as the Sun would in its entire 10-billion-year lifespan \cite{graham2013metal,gendre2013ultra}. Several satellite missions (e.g. HETE-2, FERMI, INTEGRAL, SWIFT, RXTE, ULYSSES) have been launched since the first observation of GRB and recorded flare of GRBs and provided useful data related to the origin of GRBs. The genesis and astrophysical implications of all these  processes are vehemently debated. The energy released in such processes are  estimated to be $\sim 10^{44-47}$ erg \cite{kumar2015physics,nakar2007short,berger2014short}. While the reliable nature of these sources is absolutely fascinating and not yet entirely understood, more spectacular events definitely related with them challenge the researchers' inquisitiveness. Many researchers have suggested models for $\gamma$-ray superflares, which are based on compact objects (e.g. Neutron stars, Black holes binary mergers). The model based on anisotropic fluid for the highly compact stars (i.e. solid quarks)  studied by  Xu et al. \cite{xu2006superflares} have proposed an alternative model from the conventional one (i.e. the magnetar model)\footnote{Magnetars are the most highly magnetized neutron stars in the cosmos (with magnetic field $10^{13}$–$10^{15}$ G).}, that SGR superflares could be caused by substantial quakes in  quark stars.  Xu et al. \cite{xu2006superflares} addressed the prospect of gamma-ray bursts  resulting from the phase transition of a neutron star to a quark star, with the energy released on the scale of $10^{52}$ erg. The degradation of the superstrong field ($B \geq 10^{14}$ G) in magnetars might fracture the crust and create bursts of high gamma-ray emission \cite{thompson1995soft,thompson1996soft}. Gravitational, magnetic, and superfluid forces could all exert stress on the crust of a neutron star that is evolving. Fracture of the star's crust as a result of these forces might have an impact on the star's spin dynamics and result in high-energy bursts \cite{franco2000quaking}. 

The source of the released energy is one of the most essential and fundamental problems (inferred isotropic equivalent values are given in Table \ref{tab:1}, Column 3.). There are  few options for free energy sources that are worth examining for comparison and future research.
One such attempt is related to the high-density phase transition at the core-evelope interface that can cause instabilities. Mechanical readjustment (crust-cracking) of the star crust can occur, depending on the parameters and the dynamics in which matter undergoes a phase transition which lead to potential  release of   significant amount of energy \cite{marranghello2002phase}. Based on such estimates, the amount of  energy released is limited to around $\Delta E = W (\Delta a / a)$, where $W$ signifies the star's original binding energy and $\Delta a/a$ signifies the radius's fractional change. It is estimated that energy  of the order $10^{53}$ erg is released in such processes. It has been recently improved to include an exotic solid quark phase which constitutes the bulk of the interior of the neutron star, which is believed to be a possible source of free energy \cite{xu2003solid,peng2008pulsar,zhou2004quakes,xu2006superflares}. The burning of neutron matter to strange matter may be a source of free energy. For example, if the energy per baryon number  of three-flavor quark matter is smaller than that of squeezed hadronic matter, and if other astronomical conditions are satisfied then it induces conversion of a neutron star to a strange star shortly after its formation \cite{berezhiani2002gamma,ouyed2002quark}. However, it is conceivable for a neutron star to evolve into a quark star, particularly a strange quark star. Quark stars are hypothesised extreme compact objects made of quark matter. They are widely considered as the actual ground state of compact baryon stars due to their high density. The presence of strange quark matter raises a significant condition for the selection of the equation-of-state (EoS) of quark stars: the EoS must incorporate the component provided by strange quark matter in order for the EoS to be considered credible. Strange quark matter has a lower pressure and energy than conventional quark matter with the same quark number density, permitting to be a little more stable.

\begin{tableorg}
\centering
\caption{Superflares from GRBs \cite{graham2009grb,soderberg2006afterglow} and SGRs \cite{horvath2005energetics}}
{\begin{tabular}{lll}
\hline\noalign{\smallskip}
Object & Date & $E_{\gamma}$ (erg)  \\
\noalign{\smallskip}\hline\noalign{\smallskip}
 GRB 070714b& 2007 Jul 14 & $1.2\times 10^{51}$   \\
 GRB  051221a  & 2015 Dec 21 & $(1.2-1.9)\times 10^{49}$  \\
SGR 0526-66& 1979 Mar 5 & $6\times 10^{44}$  \\
SGR 1900+14 & 1998 Aug 27 & $2 \times 10^{44}$  \\
SGR 1806-20 & 2004 Dec 27 & $3.5\times 10^{46}$  \\
\noalign{\smallskip}\hline
\end{tabular}\label{tab:1}}
\end{tableorg}

In relativistic astrophysics, the stability of a compact star structure is a major concern. Chandrasekhar \cite{chandrasekhar1964dynamical} investigated radial perturbation for isotropic fluid spheres in general relativity and proposed the pulsation equation to evaluate fluid sphere stability. Generalizing this method for use with anisotropic matter distributions was accomplished by Dev and Gleiser \cite{dev2003anisotropic}. Cracking, also known as overturning, is a notion that was first presented by  Herrera \cite{herrera1992cracking} and  Di Prisco et al.\cite{di1994tidal}, and it proposes a different approach ($-1\leq v_{s\bot}^{2}-v_{s r}^{2}\leq 1$, where $v_{s\bot}$ and $v_{s r}$ are tangential and radial speed of sound respectively)  determining whether or not an anisotropic matter distribution is stable in general relativity.  
The fundamental assumption of this approach is that the fluid constituents on each side of the cracking point experience a relative acceleration towards each other. It was developed with the purpose of describing the behavior of a fluid distribution  away from its equilibrium state. Later on, Chan et al. \cite{chan1993dynamical} showed that even very minor departures from local isotropy might contribute significantly for its structural stability. In conjunction with this, they found that changes in density by themselves do not throw the system out of equilibrium when it is configured with the anisotropic matter. Such kinds of deviations can only be induced by perturbations that affect both the density and the local anisotropy \cite{di1994tidal,di1997cracking}.

Increasing theoretical studies suggest that fascinating physical events may cause local anisotropy, such as uneven radial and tangential stresses $p_{r}\neq p_{\bot}$ ( see  \cite{herrera1997local,mak2003anisotropic}, and references therein). In the framework of General Relativity, Lemaître \cite{lemaitre1933univers} noted that local anisotropy can relax the upper limits on the maximum surface gravitational potential. Since the early works of  Bowers and Liang \cite{bowers1974anisotropic} its significance in General Relativity has been investigated.   Hillebrandt and Steinmetz  \cite{hillebrandt1976anisotropic} studied the stability of completely relativistic anisotropic neutron star models and found stability criteria comparable to that of isotropic models. Subsequently, Chan et al.  \cite{chan1993dynamical} evaluated the influence of  the local anisotropy in the emergence of dynamical instabilities. They observed that  moderate anisotropies might have a significant impact on the system's evolution.

 Many studies have been reported on core-envelope anisotropic relativistic objects in recent times \cite{sagar_2022,pant_2020three}. In which compact stars are divided into two distinct regions in accordance with the structure of various physical considerations  \cite{thomas2005core,tikekar2005relativistic,mafa2016anisotropic,gedela2019relativistic,mardan2021charged,pant_2021,gedela_2020}. Any model for an anisotropic compact object is ineffective if it is unstable against variations of its physical variables. Distinct degrees of stability or instability lead to different patterns of development in the collapse of self-gravitating objects. In this paper, we shall investigate the effect of fluctuations in density and local anisotropy have on the potential cracking of local and non-local anisotropic matter configurations within the  general relativity framework. In particular, we  concentrate on how these disturbances cause phenomena that are conducive for starquakes. Here, we incorporate Einstein's equations that determine cracking, overturning, expansion, or collapse. Certain occurrences might radically change the system's evolution. If a configuration does not really crack (or overturn), it is indeed potentially stable (not definitely stable), since further perturbations might trigger expansion or collapse.  Abreu et al. \cite{abreu2007sound}, Gonzalez et al. \cite{gonzalez2015cracking}, and Ratanpal \cite{ratanpal2020cracking}
proved that the regions for which $-1\leq v_{s\bot}^{2}-v_{sr}^{2}\leq 0$ are potentially stable and the regions for which $0\leq v_{s\bot}^{2}-v_{sr}^{2}\leq 1$ are potentially unstable , where $v_{s\bot}^{2}=\frac{\partial p_\bot}{\partial \rho}$ and $v_{sr}^{2}=\frac{\partial p_{r}}{\partial \rho}$.


Thus, compact stars are astrophysical laboratories of many extreme  physics. The core and envelope of a highly compact stars are made up of distinct physical materials, according to the current knowledge of the strong interactions processes leading to a phase transition to quark matter, the study becomes extremely difficult due to the non-perturbative aspects of quantum chromodynamics (QCD)\cite{annala2020evidence}. An alternative treatment to such a self bound system is being studied recently \cite{khunt2021distinct} based on geometrical approach wherein the core-envelope model has been employed. The model with isotropic core and anistropic envelope naturally supports quake formation.
Following such studies  \cite{xu2006superflares} and  \cite{shu2017gamma}, the prospects of  a progenitor for   starquakes  from a self bound compact stars are being addressed in this study. 
The core-envelope models studied by Thomas, Ratanpal and Vinodkumar (TRV model) have considered anisotropic pressure in the envelope region and isotropic pressure in the core region \cite{thomas2005core}.  Physically such a scenario is  possible with a core containing  pure quark phase and the envelope accumulating quark-hadron mixed phase. The comprehensive investigation of the TRV model is described in our previous work \cite{thomas2005core,khunt2021distinct}, the structural properties like mass-radius relation, gravitational red-shift, Keplerian frequency, and surface gravity are studied using the TRV model. Their structural properties show that this model is important for the study of highly compact self-bound stars. A density perturbation at the core-envelope interface can lead to an event like starquake. The rearrangement of the star's mass distribution caused by a starquake affects the star's moment of inertia, resulting in precession and polar drifting \cite{link1998starquake}. As a result, unusual spin behaviour might be an indicator of the occurrence of a stellar quake. Evidence of crust cracking may already exist in few isolated pulsars, showing that the process is underway.
  
We utilize the core-envelope model based  on the geometrical approach of highly compact self bound stars to compute the stress-energy stored due to the pressure anisotropy at the core-envelope interface that would be released in the event of a quake like scenario.  In Sec.\ref{sec:2}, we  present anisotropic Tolman–Oppenheimer–Volkoff equations, and we describe briefly the TRV model and express the pressure anisotropy parameter $(S)$ at the interface of the core and envelope. 
In Sec. \ref{sec:3}, we discuss general relativity inspired EoS and stellar stability.  We then investigate the impact of possible anistropy in the stability and cracking in Sec.\ref{sec:4}.  In Sec.\ref{sec:5}, we  estimate the amount of radius, gravitational energy and moment of inertia due to change of anisotropy. Finally, we summarize our main findings in Sec. \ref{sec:6} and conclude in Sec. \ref{conclusion}.

\section{Anisotropic matter configuration in general relativity}
\label{sec:2}

The following spherically symmetric line element in the Schwarzchild coordinates $(x^{i})$=$(t, r, \theta, \phi)$ describes the interior of an anisotropic fluid sphere. We start therefore with a metric of the form 
\\
\begin{equation} 
    ds^{2} = e^{\nu (r)} dt^{2}- e^{\lambda(r)} dr^{2}- r^{2} d\theta^{2}- r^{2} \sin^{2} \theta d\phi^{2},
    \label{sch}
\end{equation}
where $\nu$ and $\lambda$ are functions of the radial coordinate $(r) $ only,  and a fluid stress-energy tensor $T^{\mu}_{\nu} = $ diag $[\rho,- p_{r}, -p_{\bot}, -p_{\bot}]$, where $p_{r}$ and $p_{\bot}$ are the radial and tangential pressures, respectively. The metric's proper boundary condition has to be  matched  with the Schwarchild exterior metric on the star's surface. It's done as follows

\begin{equation} \label{sch_ext}
    ds^{2} =
     \bigg(1-\frac{2 G M}{a c^{2}}\bigg) dt^{2}- \bigg(1-\frac{2G M}{a c^{2}}\bigg)^{-1} dr^{2} 
   - r^{2} d\theta^{2}- r^{2} \sin^{2} \theta d\phi^{2} 
\end{equation}

\begin{equation}
    \nu(r=a)=\ln \bigg(1-\frac{2GM}{a c^{2}}\bigg)
\end{equation}

\begin{equation}
    \lambda(r=a)=-\ln \bigg(1-\frac{2GM}{a c^{2}}\bigg).
\end{equation}
 The radius and mass of the star are represented by $a$ and $M$, respectively.
 
  The Einstein field equation is given by \cite{weinberg1972gravitation}
\begin{equation}\label{einstein}
    \mathcal{R}_{\mu \nu}-\frac{1}{2} \mathcal{R} g_{\mu \nu }=- \frac{8\pi G}{c^{4}} T_{\mu \nu}
\end{equation}
for an energy–momentum tensor relevant for ideal fluid has been solved for the metric provided by Eq.~(\ref{sch}).
The  energy-momentum tensor for anisotropic fluid distribution is taken as

 \begin{equation}
    T_{\mu \nu}=(\rho +p)u_{\mu}u_{\nu}-p g_{\mu \nu} +\pi_{\mu \nu}
    \label{anis}
\end{equation}
 where  $\pi_{\mu \nu}$ denotes anistropic stress tensor given by \cite{thomas2005core}
 
 \begin{equation}
     \pi_{\mu \nu}=\sqrt{3} S \bigg[C_{\mu}C_{\nu}-\frac{1}{3}(u_{\mu}u_{\nu}-g_{\mu \nu})\bigg]
 \end{equation}
where $S$=$S(r)$ is the magnitude of anisotropy stress tensor and $C^{\mu}=(0,  -e^{-\frac{\lambda}{2}},0,0)$, which is a radial vector.

The Einstein filed equations for this spacetime geometry and matter distribution are

\begin{equation}
    \frac{(1-e^{-\lambda})}{r^{2}}+\frac{\lambda^{'} e^{-\lambda}}{r}= 8\pi \rho ,
    \label{density}
\end{equation}

\begin{equation}
    \frac{\nu^{'} e^{-\lambda}}{r}-\frac{(1-e^{-\lambda})}{r^{2}}= 8\pi P_{r} ,
    \label{eife}
\end{equation}

\begin{equation}
    \frac{e^{-\lambda}}{4} \Big(2 \nu^{''}+\nu^{' 2}-\nu^{'}\lambda^{'}+\frac{2\nu^{'}}{r}-\frac{2 \lambda^{'}}{r}  \Big)= 8\pi P_{\bot} .
    \label{eafe}
\end{equation}
where  prime denotes  here differentiation with respect to the radial coordinate $r$.

Using Eqs. (\ref{eife}) and (\ref{eafe}), or correspondingly the conservation law $T^{\mu}_{\nu ;\mu}=0$, it is now convenient to transform the above equations into a form where the hydrodynamical properties of the system are more evident and that reduces to the TOV equations for systems with isotropic pressure, i.e.,

\begin{equation}
    e^{-\lambda}=1-\frac{2 m(r)}{r},
\end{equation}

\begin{equation}
    \nu^{'}=\frac{2m(r)+8\pi r^{3}P_{r}}{r(r-2m(r))},
    \label{nuu}
\end{equation}

\begin{equation}
    P^{'}_{r}=-(\rho +P_{r})\frac{\nu^{'}}{2} +\frac{2}{r}(P_{\bot}-P_{r}),
    \label{rpre}
\end{equation}
where the \textit{mass function} $m(r)$ is defined by 
\begin{equation} \label{mass}
m(r)=4\pi \int_{0}^{r} \rho ~ \bar r^2 d\bar r.
\end{equation}
Moreover, it corresponds to the mass inside  a sphere of  radius $r$ as perceived by a distant observer.
 
 Combining (\ref{nuu}) and (\ref{rpre}), we finally obtain
 
 \begin{equation}
     P^{'}_{r}=-(\rho+P_{r})\frac{m(r)+4\pi r^{3} P_{r}}{r(r-2m(r))}+\frac{2{\Delta}}{r}P_{r}.
     \label{hydro}
 \end{equation}
 which is the generalized Tolmann-Oppenheimer-Volkoff (TOV) equation. It can be seen that Eq. (\ref{hydro}) readily reduces to the standard TOV when $P_{r}=P_{\bot}$. Where $P_{\bot}=(1+\Delta)P_{r}$ is introduced here to account for the anisotropy in the envelope.  Interestingly, Eq.~(\ref{hydro}) which recalls the Newtonian hydrostatic-equilibrium equation and where the final term is obviously zero in the case of isotropic pressures, is evocative of the Newtonian hydrostatic equilibrium equation, i.e., $P_{\bot}=P_{r}$.\\

\subsection{The TRV core-envelope model}

The model in the core-envelope family, as discussed by  Thomas et al.  \cite{thomas2005core}, is our primary focus in this paper.
To solve Einstein's equations, they used ansatz for a pseudospheroidal geometry of spacetime. The potential for this spacetime geometry is expressed as \cite{thomas2005core}.

\begin{equation} \label{pot}
    e^{\lambda(r)}=\frac{1+K \frac{r^{2}}{R^{2}}}{1+\frac{r^{2}}{R^{2}}}
\end{equation}\\
\textit{K} and \textit{R} are geometric variables, respectively.
With anisotropic stress tensor $\pi_{\mu\nu}$, the energy–momentum tensor components (\ref{anis}) have nonvanishing components.
\begin{equation} \label{enemom}
    T^{0}_{0}=\rho,\,\,\ T^{1}_{1}=-\bigg(p+\frac{2S}{\sqrt{3}}\bigg),\,\,\ T^{2}_{2}=T^{3}_{3}=-\bigg(p-\frac{p}{\sqrt{3}}\bigg).
\end{equation}
The magnitude of anisotropic stress is calculated using the following equation
\begin{equation}
     S=\frac{\mid p_{r}-p_{\perp}\mid}{\sqrt{3}}
 \end{equation}
The core and envelope region's boundary conditions are as follows
\begin{equation}\label{condition}
  \begin{split}
  S(r) =0 \,\,\mbox{for}\,\,\,\, 0\leq r \le R_{C}, \\
  S(r)\neq 0\,\,\mbox{for}\,\,\,\, R_{C} < r \le R_{E}
  \end{split}
\end{equation}
where $R_{C}$ represents the core boundary radius and $R_{E}$ represents the envelope boundary radius which is same as the radius of the star $(a)$ under consideration. Using the conditions (\ref{condition}) with 
Eqs. (\ref{sch}), (\ref{einstein}), (\ref{pot}) and (\ref{enemom}), the equations for pressures are given by Einstein's field equations.

The Einstein field equation(\ref{einstein})  relating to the metric (\ref{sch}) employing ansatz (\ref{pot}) is provided by a combination of three equations:

\begin{equation}
    8\pi\rho = \frac{K-1}{R^2} \bigg(3+ K\frac{r^2}{R^2}\bigg) \bigg(3+ K\frac{r^2}{R^2}\bigg)^{-2},
\end{equation}

\begin{equation}
    8\pi p_{r} = \bigg[\bigg( 1+ \frac{r^2}{R^2}\bigg) \frac{\nu^{'}}{r}-\frac{K-1}{R^2}\bigg] \bigg(1+K \frac{r^{2}}{R^{2}}^{-1}\bigg),
\end{equation}

\begin{equation}\label{ani_eq}
\begin{split}
8\pi \sqrt{3}S 
&=-\bigg(\frac{\nu^{''}}{2}+\frac{\nu^{'2}}{4}-\frac{\nu^{'}}{2r}\bigg) \bigg(1+\frac{r^{2}}{R^{2}}\bigg) \bigg(1+K \frac{r^{2}}{R^{2}}\bigg)^{-1}+   \\
& \frac{(K-1)}{R^{2}}r \bigg(\frac{\nu '}{2}+\frac{1}{r}\bigg)  \bigg(1+K \frac{r^{2}}{R^{2}}\bigg)^{-2}+  \\
& \frac{K-1}{R^{2}}         \bigg(1+K \frac{r^{2}}{R^{2}}\bigg)^{-1}.
\end{split}
\end{equation} 

Thomas et al. \cite{thomas2005core} considered a case  with isotropic core and anisotropic envelope with radial pressure $p_{r}$ and tangential pressure $p_{\bot}$.  The anisotropy begins forming from the core boundary has radius $ r = R_{C}$. The radial difference in pressure in the enveloping region and it reduces to zero at the surface ($ r = a$, where an is the radius of the star under consideration). We present the core upto the radius $r=R_{C}$, during which $S(r) = 0$. The radius of the star is considered as a and we split it into two parts as conditions present in Eq.(\ref{condition}).

\subsubsection{The Core of the star}

The isotropic distribution of matter defines the core of the compact star. The radial pressure $p_{r}$ is thus equal to the tangential pressure $p_{\bot}$ across the core region $0\leq r \leq R_{C}$, and $S(r) = 0$. Eq. (\ref{ani_eq}) thus becomes

\begin{equation} \label{non_liner}
\begin{split}
&\bigg(\frac{\nu^{''}}{2}+\frac{\nu^{'2}}{4}-\frac{\nu^{'}}{2r}\bigg) \bigg(1+\frac{r^{2}}{R^{2}}\bigg) \bigg(1+K \frac{r^{2}}{R^{2}}\bigg)^{-1}-  \\
& \frac{(K-1)}{R^{2}}r \bigg(\frac{\nu '}{2}+\frac{1}{r}\bigg)  \bigg(1+K \frac{r^{2}}{R^{2}}\bigg)^{-2}=0
\end{split}
\end{equation} 

\noindent Equation (\ref{non_liner}) is a non-linear differential equation, if we adopt new independent variable $z$ and dependent variable $ F$ described by:
\begin{equation}
    z=\sqrt{1+\frac{r^{2}}{R^{2}}},
\end{equation}
\begin{equation}
    F=e^{\nu/2},
\end{equation}
Equation (\ref{non_liner}) has the linear form

\begin{equation}
    (1-K+Kz^{2}) \frac{\mathrm{d}^2F}{\mathrm{d}z^{2}}- Kz\frac{\mathrm{d}F}{\mathrm{d}z}+K(K-1)F=0.
\end{equation}

\noindent We get the solution of the metric potential ($e^{\nu/2}$) by transforming this differential equation and solving it, we get
\begin{equation}
    F= e^{\nu/2}= A\sqrt{1+\frac{r^{2}}{R^{2}}}+B \bigg[\sqrt{1+\frac{r^{2}}{R^{2}}}L(r)+\frac{1}{\sqrt{2}}\sqrt{1+2\frac{r^{2}}{R^{2}}}\bigg]
\end{equation}

\noindent As a result, the spacetime metric of the core region $0\leq r \leq R_{C}$ is described as follows:

\begin{equation}
\begin{split}
\mathrm{d}s^{2}
 &=\bigg( A\sqrt{1+\frac{r^{2}}{R^{2}}}+B \bigg[\sqrt{1+\frac{r^{2}}{R^{2}}}L(r)+\frac{1}{\sqrt{2}}\sqrt{1+2\frac{r^{2}}{R^{2}}}\bigg]\bigg)^{2} dt^{2} + \\
 & \bigg(\frac{1+2 \frac{r^{2}}{R^{2}}}{1+\frac{r^{2}}{R^{2}}}\bigg) dr^{2}-r^{2}d\theta^{2}-r^{2}\mathrm{sin}\theta d\phi^{2}.
    \end{split}
\end{equation}

\noindent The pressure and metric coefficients must be continuous across the core boundary, $r = R_{C}$, in order to determine the constants $A$ and $B$. Accordingly, the density distribution (core and envelope region) is expressed as 
\begin{equation}\label{den}
    \rho = \frac{1}{8\pi R^{2}}\bigg[ 3+2\frac{r^{2}}{R^{2}}\bigg] \bigg[1+2 \frac{r^{2}}{R^{2}}\bigg]^{-2}.
\end{equation}
\\
 where \textit{R} is a geometrical parameter. Eq.~(\ref{den}) provides the density distribution in core and envelope region by using boundary condition for $0\leq r\leq R_{C}$ for core and $ R_{C} \leq r\leq R_{E}$ for envelope region.

As a result, the pressure distribution (both core and envelope region) is expressed as :


\begin{align}\label{corepre}
8\pi  p_{\scaleto{C}{3pt} }=  \frac{A\sqrt{1+\frac{r^{2}}{R^{2}}}+B \bigg[\sqrt{1+\frac{r^{2}}{R^{2}}}L(r)+\frac{1}{\sqrt{2}}\sqrt{1+2\frac{r^{2}}{R^{2}}}\bigg]}{ R^{2} \big(1+2\frac{r^{2}}  { R^{2}}\big) \bigg [A +\sqrt{1+2\frac{r^{2}}{R^{2}}}+B \bigg(\sqrt{1+\frac{r^{2}}{R^{2}}}  L(r) -\frac{1}{\sqrt{2}} \sqrt{1+ 2 \frac{r^{2}}{R^{2}}}\bigg) \bigg]}
\end{align}
for the radial pressure in the core region.

  Where
\begin{equation}
    L(r)= \ln\bigg(\sqrt{2}\sqrt{1+\frac{r^{2}}{R^{2}}} +\sqrt{1+2 \frac{r^{2}}{R^{2}}}\bigg).
\end{equation}

 The constants $A$ and $B$ are to be determined by requiring that the pressure and metric coefficient must be continuous  across the core boundary $r=R_{C}$; and this is done as follows :
The anisotropy parameter vanishes at the core boundary, when $r = R_{E}$. Therefore, $\frac{a^2}{R^2} >2 $ is necessary due to the tangential pressure's positive sign. The consistency of metric coefficients and pressure across the distribution at $ r = R_{C}$ leads to
 \begin{equation}
 \sqrt{3} A + B \left[ \sqrt{3} L(R_{C})+ \sqrt{2.5}  \right]= 5^{-3/4} \left[11 \sqrt{3}C+D   \right],
 \end{equation}
\begin{equation}
 \sqrt{3} A + B \left[ \sqrt{3} L(R_{C})+ \sqrt{2.5}  \right]= 5^{1/4} \left[11 \sqrt{3}C+D   \right],
 \end{equation}
 where 
  \begin{equation}
      L(R_{C}) = \text{ln} (\sqrt{5}+\sqrt{6}).
  \end{equation}

\begin{align}
    A=\frac{[5\sqrt{5}- 3\sqrt{2} (\sqrt{3} L(R_{c})-\sqrt 2.5)   ] C+ \frac{1}{\sqrt{3}} [5 \sqrt{5}+2\sqrt{2}(\sqrt{3} L(R_{c})-\sqrt{2.5}     ]      D       }
    {5\frac{5}{4}},
 \end{align}

\begin{equation}
    B= \frac{ \sqrt{2} } {5\frac{5}{4}} [3\sqrt{3}C-2D].
\end{equation}
The curvature parameter $R$ is given by 
\begin{alignat}{2}
 R=\sqrt{\frac{3\lambda}{8\pi \rho(a)}} &\quad , \:  \lambda=\frac{\rho(a)}{\rho(0)}= \frac{1+\frac{2 a^{2}}{3 R^{2}}}
  {(1+2\frac{a^{2}}{R^{2}})^{2}}.
\end{alignat}
\noindent where $\lambda$ is density variation parameter.\\
\\

\subsubsection{The Envelope of the Star} \label{envelope_sec}

The envelope of the star is defined by the anisotropic concentration of matter. Thus all across envelop region $R_{C} \leq r \leq R_{E}$ the radial pressure $p_{r}$ is different from the tangential pressure $p_{\bot}$, and therefore $ S(r)\neq 0$. To determine the solution of Eq. (\ref{ani_eq}), in this context, we introduce new variables $z$ and $\psi$ specified by:

\begin{equation} \label{anzat_2}
    z= \sqrt{1+\frac{r^{2}}{R^{2}}} \;\;\;\;\;\;\;\;   \psi = \frac{e^{\nu/2}}  {(1-K+Kz^{2})^{1/4}}
\end{equation}

\noindent in terms of which Eq. (\ref{ani_eq}) has the following form:
\begin{equation} \label{diff_eq_2}
    \frac{d^{2}\psi}{dz^{2}}+\bigg[\frac{2K(2K-1)(1-K+Kz^{2})-5K^{2}z^{2}}   {4(1-K+Kz^{2})^{2}} + \frac{8 \sqrt{3}\pi R^{2}S(1-K-Kz^{2})}  {z^{2}-1} \bigg]\psi=0
\end{equation}
\noindent On prescribing 
\begin{equation}
    8\pi \sqrt{3}S =- \frac{(z^{2}-1) [2K(2K-1)(1-K+Kz^{2})-5K^{2}z^{2}]} {4R^{2}(1-K+Kz^{2})^{3}}
\end{equation}

\noindent the second term in  Eq. (\ref{diff_eq_2}) eliminates and the resultant equation is
\begin{equation}
\frac{d^{2}\psi}{dz^{2}} =0,
\end{equation}

\noindent It has solution of the form :

\begin{equation}
    \psi = Cz+D
\end{equation}
where $C$ and $D$ are constants of integration. From Eq. (\ref{anzat_2}) we obtain  

\begin{equation}
e^{\nu/2}= \bigg(1+K\frac{r^{2}}{R^{2}}\bigg)^{1/4}  \bigg(C\sqrt{1+\frac{r^{2}}{R^{2}}}+D\bigg).  
\end{equation}

\noindent Thereby the spacetime metric of the enveloping region $b\leq r \leq R_{E}$   is described by:

\begin{equation}
\begin{split}
\mathrm{d}s^{2}
 &=\sqrt{1+K\frac{r^{2}}{R^{2}}} \bigg(C\sqrt{1+\frac{r^{2}}{R^{2}}}+D\bigg)^{2} dt^{2} +
 \bigg(\frac{1+K\frac{r^{2}}{R^{2}}}{1+\frac{r^{2}}{R^{2}}}\bigg) dr^{2} \\
& -r^{2}d\theta^{2}-r^{2}\mathrm{sin}\theta d\phi^{2}.
    \end{split}
\end{equation}

\noindent We shall set $K=2$ such that the distribution follows the same 3-space geometry.

\noindent The following are explicit formula for the radial pressure $p_{r}$, tangential pressure $p_{\bot}$, and anisotropy parameter $S(r)$:

\begin{equation}\label{envop1}
    8\pi p_{\scaleto{E}{3pt}}= \frac{C\sqrt{1+\frac{r^{2}}{R^{2}}}\big(3+4\frac{r^{2}}{R^{2}}\big)+D} 
    {R^{2} \big(1+2\frac{r^{2}}{R^{2}}\big)^{2}\big(C \sqrt{1+\frac{r^{2}}{R^{2}}}+D\big)},
\end{equation}

\begin{equation}\label{envop2}
    8\pi p_{\scaleto{E}{3pt}\bot}=   8\pi p_{\scaleto{E}{3pt}}
   -\frac{\frac{r^{2}}{R^{2}}\big(2-\frac{r^{2}}{R^{2}}\big)}
  {R^{2} \big(1+2\frac{r^{2}}{R^{2}}\big)^{3}},
\end{equation}
 and anisotropy $S$ is expressed as
 
  \begin{equation}\label{anisotropy}
  8\pi\sqrt{3} S= \frac{ \frac{r^{2}}{R^{2}}\big(2-\frac{r^{2}}{R^{2}}\big)  }
  {R^{2}\big(1+2 \frac{r^{2}}{R^{2}}\big)^{3}}.
 \end{equation}
\noindent The constants $C$ and $D$ are to be determined by matching the solution with Schwarzchild exterior solution (\ref{sch_ext}). The following relationships are implied by the continuity of metric coefficients and pressure along radial direction across $r = a$:

\begin{equation} \label{48}
    e^{\nu(a)} =  \frac{1+ \frac{a^2}{R^2}}{1+2 \frac{a^2}{R^2}} = 1-\frac{2M}{a}
\end{equation} 

\begin{equation} \label{49}
    C \sqrt{1+\frac{a^2}{R^2}} \left( 3+4 \frac{a^2}{R^2}\right)+D=0.
\end{equation}

\noindent Eqs.(\ref{48}) and (\ref{49}) determine constants  $C$ and $D$ as 
\begin{equation}
C=-\frac{1}{2}\bigg( 1+2\frac{a^{2}}{R^{2}}\bigg)^{-\frac{7}{4}},
\end{equation}

\begin{equation}
    D=\frac{1}{2}  \sqrt{1+\frac{a^{2}}{R^{2}}} \bigg(3+4 \frac{a^{2}}{R^{2}}\bigg) \bigg(1+2 \frac{a^{2}}{R^{2}}\bigg)^{-\frac{7}{4}}.
\end{equation}

According to Eq. (\ref{den}), the matter density at the centre is explicitly related to the curvature parameter $R$ as

\begin{equation} \label{52}
   \lim_{r\to 0} \rho (r) = \rho(0) ; \;\; \therefore \;\; 8\pi \rho (0)= \frac{3}{R^2}
\end{equation}

\noindent Eq.(\ref{52}) estimates $R$ in terms of $\rho(a)$ and $\lambda$.

\begin{figure}[H]
\centerline{\includegraphics[scale=0.4]{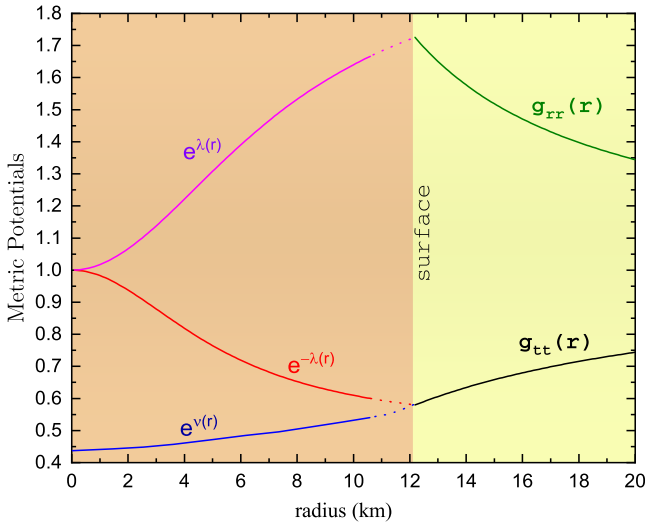}}
\vspace*{1pt}
\caption{(Color online) Metric function of static star as function of radius in units of km. For a $\lambda=0.07$ . The redshift factor $e^{\nu(r)}$ steadily increases from the center of the star towards the asymptotic region. The metric function $e^{-\lambda(r)}$ is flat near the center and reaches a minimum near the surface of the star, where it joins the redshift factor.
 \label{f11}}
\end{figure}

Fig. \ref{f11} depicts the radial profile of the temporal and spatial metric potentials, $e^{\nu(r)}$ and $e^{-\lambda(r)}$, respectively. Both the metric potentials are clearly finite at their centres and maintain consistently at all points. The redshift factor $e^{\nu(r)}$ steadily increases from the center of the star towards the asymptotic region. The metric function $e^{-\lambda(r)}$ is flat near the center and reaches a minimum near the surface of the star, where it joins the redshift factor. The metric potentials $e^{-\lambda(r)}$, $e^{\nu(r)}$  are continuous and found well behaved in the core and the envelope regions.

\begin{figure}[h]
\centerline{\includegraphics[scale=0.35]{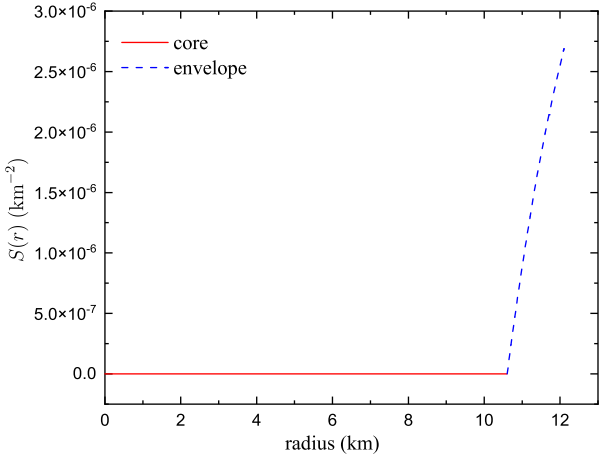}}
\caption{(Color online) Variation of an anisotropy $S$ in km$^{-2}$ with respect to a radius of the star within a range of boundary condition as given in Eq.~(\ref{condition}). For a density variation $(\lambda=0.07)$.} 
\label{fig:1}       
\end{figure}

The radial and tangential pressures differ due to local pressure anisotropy ($p_{r}\neq p_{\perp}$). In a conventional isotropic Tolman–Oppenheimer–Volkoff  equation, the difference between radial and tangential pressure generates an additional force. The discussion on anisotropic pressure and the physical conditions for anisotropic stars can be found in Refs. \cite{sulaksono2015anisotropic, setiawan2019anisotropic}, and further details can be seen
in an anisotropic pressure review paper \cite{herrera1997local}. In the present work,
we  examine quantitatively, the changes on the properties of the compact star due to  the local anisotropic pressure. In order to adopt the model suggested in \cite{thomas2005core}, we have computed anisotropic magnitude for the core-envelope regions using the conditions (\ref{condition}) with Eqs.~(\ref{corepre}), (\ref{envop1}), (\ref{envop2}) and (\ref{anisotropy}). For a density variation $(\lambda=0.07)$, the distributed anisotropy for the core and the envelope regions is shown Fig. \ref{fig:1}.
For the core region, the anisotropy $(S)$ is zero according to the TRV model \cite{thomas2005core}, while in the envelope region anisotropy is non-zero. The anisotropy magnitude inside the envelope region computed using the Eq.~(\ref{anisotropy}) and their result is represented as blue solid curve in  Fig.\ref{fig:1}. It can be noticed that the anisotropy is an increasing function of $r$. For 10.617 km $\leq r$ $\leq$ 12.177 km, the anisotropy varies in the range  $[2.458 \times 10^{-8}, 2.774\times 10^{-6}]$. It is  important to note its influence on the stability of the envelope region. The main reason for choosing a thin crust is to study processes like cracking or overturning in the medium of anisotropy. To avoid confusion, we provide the anisotropy profile for various density variation parameter, $\lambda$ in  Appendix A (\ref{app_1}).

\subsection{Revisited physical plausibility}
The interior solution should fulfill certain fundamental physical criteria. Some of the physical acceptability conditions for anisotropic materials have been described in \cite{herrera1997local,thomas2005core,mak2003anisotropic} as

\begin{enumerate}[label=(\roman*)]
\item radial pressure $p_{r}$, tangential pressure $p_{\bot}$ and density $\rho$  should be positive everywhere within the configuration;
\item gradients for radial pressure and density should be negative, 
\begin{equation*}
  \frac{\partial p_{r}}{\partial r} \leq 0   \;\;\; \text{and} \;\;\;  \frac{\partial \rho}{\partial r} \leq 0
\end{equation*}

\item Sound speed at the interface should be continuous and should satisfy the causality condition at the core of a compact star model, as well as decreasing monotonically outwards.
\begin{equation*}
    \frac{\partial p_{r}}{\partial \rho} \leq 1   \;\;\; \text{and} \;\;\;  \frac{\partial p_{\bot}}{\partial \rho} \leq 1
\end{equation*}

\item  The core and the envelope for the star should satisfy the energy conditions besides being
continuous at the interface.
\end{enumerate}

\section{The geometrical EoS of extreme dense matter and stellar stability }
\label{sec:3}

In this work, we use geometrically deduced EoSs which are based on our previous work \cite{khunt2021distinct} to study the canonical properties of compact stars. In which, the pressure and density profile are computed using the Eqs.~(\ref{den}), (\ref{corepre}) and (\ref{envop1}) for different choices density parameter ($\lambda=0.01$ and $\lambda =0.07$). The best fit for the pressure-density curve is found to be in the quadratic form 

\begin{equation}
   p=\gamma+\alpha\rho+\beta\rho^{2}
\end{equation}
where  $\gamma$, $\alpha  $  and $\beta$ are the fitted parameters. For two different density variation parameters and their fitted value of EoSs are represented in Table {\ref{tab:2}}. Based on such an equation of state, the mass-radius relation has been reported in our earlier study \cite{khunt2021distinct}. The TRV model  based upon pseudo-spheroidal geometry,  is also valid for different choice of $\lambda$. Thomas et al. \cite{thomas2005core} shows that $\lambda$ can varies from $0.01$ to $0.09$ for this particular model of superdense stars. Looking for a thin crust situation, we  present our results for the choice of $\lambda=0.07$.  The results of core radius  and envelope size ($R_{E}=a-R_{c}$)  for two different values of $\lambda$  can be viewed from Table {\ref{tab:2}}.

\begin{tableorg}[h]
 \centering
\caption{Values of constants that generate EoSs, core and envelope radii ($R_{C}$, $R_{E}$) for two different values of density variation parameters}
\scalebox{0.90}{%
{\begin{tabular}{cccccc}
\hline\noalign{\smallskip}
$\lambda$ & $\gamma$ & $\alpha$  & $\beta$ & \makecell{$R_{C}$ \\(km)}  & \makecell{$R_{E}$ \\(km)}  \\
\noalign{\smallskip}\hline\noalign{\smallskip}
 0.01& $-9.30 \times 10^{-4}$ & 1.69  &   406 & 4.00  & 7.72\\
 0.07 & $-2.09\times 10^{-4}$ & $0.77$  &$1291.49$ & 10.60 &1.57 \\
\noalign{\smallskip}\hline
\end{tabular}\label{tab:2}}}
\end{tableorg}

\begin{figure}[h]
\centerline{\includegraphics[scale=0.40]{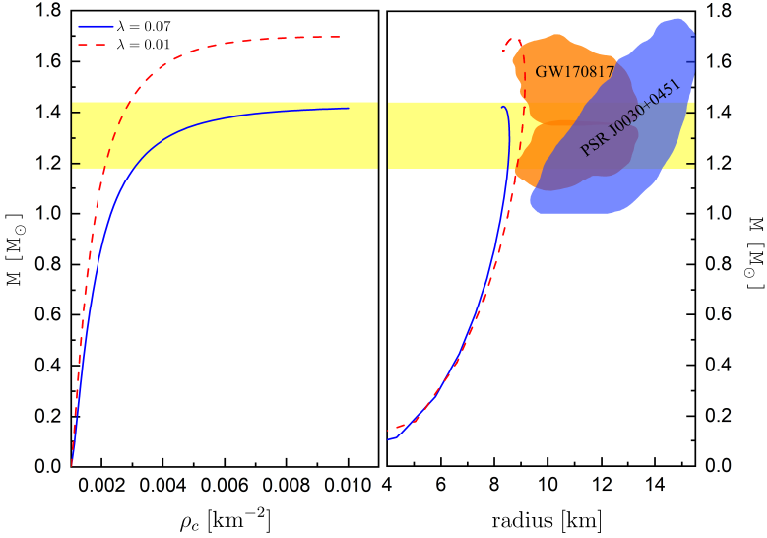}}
\caption{(Color online) Mass–central  density relationship  (panel (a))   and Mass–radius relationship (panel (b)) for the two geometrical EoSs models
considered in these plots for comparison. The hatched (Yellow color) region in panel (a) and (b) represents the range of precisely measured masses of binary radio pulsar \cite{lattimer2012nuclear}. The  orange regions are the mass–radius constraints from the GW170817 event \cite{abbott2018gw170817}. The blue region represent the pulsar, is the NICER estimations of PSR J0030+0451 \cite{miller2019psr}.}
\label{fig:mr}       
\end{figure}

In Fig.\ref{fig:mr}, we show the Mass-Radius diagram with the TRV EoS considered in this work. The density profile at a stellar structure's centre, $r = 0$,  determines how stable \footnote{Not all branches of sequence $M=M(\rho_{c})$ are stable. This can be unstable by means of radial oscillations. Degenerate stars with $dM/d\rho_{c}<0$ are found to be unstable and will finally collapse towards Neutron stars, or Black holes.} it is against its own gravitational attraction. At an equilibrium radius, the confined mass will be larger as  the  centre density is higher. The mass of a self - gravitating compact star increases with central density, in the stable region, in accordance with the Harrison-Zel'dovich-Novikov criterion \cite{zeldovich_1971,harrison1_965,saklany_2023}.   The static stability condition necessitates that $dM/d\rho_{c}>0$ for all points within the confining region, where $r < a$. As can be seen from Fig.\ref{fig:mr} (panel (a)), the mass of the model star $M({\rho_{c}})$ increases monotonically with $dM/d\rho_{c}>0$, resulting to a stable stellar structure as the central density $\rho_{c}$ increases.  As a consequence, we find by  solving the TOV equation (Eq.(\ref{hydro}) with $\Delta=0$) for a given EoS a maximum possible mass  corresponding to the central maximal density. Mass-radius decreases with decreasing density variation parameter $\lambda$.  The maximum mass at maximum central density is about $1.68 M_{\sun} $ and the corresponding  stellar radius 8.76 km for $\lambda=0.01$, while for $\lambda=0.07$ the stellar mass is about $1.42M_{\sun}$ and radius 8.33 km. The TRV model of superdense stars is  very useful for classification of different class of compact stars. For stars with $M \geq \text{\(M_\odot\)}$, the radius changes very little with $M$, $a \simeq 8-10$ km, comparable to neutron stars with a moderately stiff EoS. However, at smaller $M$ the radius of bare strange star behaves in an entirely different manner. It falls monotonically as $M$ decreases, with $a \propto M^{1/3} $ for $M \leq 0.5\text{\(M_\odot\)}$. as expected from the fact that low-mass strange stars can be described by the Newtonian theory, which gives $M \simeq \frac{4\pi}{3} \rho a^{3}$ . The decrease of $a$ with decreasing $M$ is unique feature of strange star and quark stars or compact star built of abnormal matter. It's important to note that the mass-radius constraint from the gravitational wave event GW170817 represented by the orange clouded region corresponding to the heavier and lighter neutron star, respectively \cite{abbott2018gw170817} (see insert in the Fig. \ref{fig:mr}. In this figure,  the blue clouded region show the NICER (Neutron Star Interior Composition Explorer) mass-radius measurements on  PSR J0030+0451  \cite{miller2019psr,riley2019nicer}. Besides, the yellow horizontal band show the range of precisely measured masses of binary radio pulsar \cite{lattimer2012nuclear}. Because we have the results of various EoSs in GR, we use the geometrically deduced EoS to compute the mass-radius relation for different density variation parameters, $\lambda$ that matches  with the observational data. Strictly speaking we investigated the mass-radius relation of neutron star in a general relativistic framework for the class of highly compact self-bound stars (e.g,. quark stars).

After calculating an equilibrium stellar model we analyze its stability as stable equilibrium are of astrophysical relevance. Hereafter we briefly address the stability with respect to density perturbations. For simplicity, we restrict ourselves to non-rotating spherically symmetric equilibrium models.

\section{Cracking  Stability}
\label{sec:4}

 The sound speed is a microscopic quantity of immense importance, particularly the potential upper bounds that might be obtained, because it impacts the EoS. One approach to quantify dense materials is by the velocity of sound, which is given by $v_{s}^{2}= dp/d\rho$, where $p$ is the pressure and the $\rho$ matter density. Causality demands an absolute restriction on $v_{s}\leq 1$, and thermodynamic stability ensures that $v_{s}^2 \geq 0$ \cite{lattimer2001neutron}. The fundamental assumption concerning the upper bound of the velocity of sound is that it satisfies causality ($v_{s}^{2}= dp/d\rho \leq c^2$), i.e. the speed of sound cannot reach the speed of light, which is the significant barrier imposed on the speed of sound  by general principles according to Zel’dovich et al. \cite{YBZ}. Moreover,  Bedaque and Steiner \cite{bedaque2015sound} showed that the possibility of neutron stars with masses of around two solar masses, together with knowledge of the EoS of hadronic matter at low densities, is incompatible with the constraint $c / \sqrt{3}$. The speed of sound and its impact on tidal deformability have also been analyzed by many authors \cite{reed2020large,moustakidis2017bounds,van2017upper,ma2019sound}.

When examining arbitrary and independent density and anisotropy perturbations (as in all earlier cracking investigations \cite{herrera1992cracking,di1994tidal,di1997cracking,abreu2007sound,ratanpal2020cracking}), there are few physical criteria to determine the magnitude (absolute and/or relative) of the perturbation, i.e., relatively small (or huge) the perturbations should be a cracking-like scenario could be produced by perturbations of varying orders of magnitude (and relative size $\delta \Delta / \delta \rho$). In addition, these earlier works have only considered continuous perturbations in their studies. It is possible that variable perturbations have the potential to be more effective in inducing cracking within a specific matter configuration. However, we are looking for physical characteristics that can be tested to see whether cracks are forming.
\begin{equation} \label{pertub}
    \frac{\delta \Delta}{\delta \rho} \sim \frac{\delta (p_{\bot}-p_{r})}{\delta \rho} \sim \frac{\delta p_{\bot}}{\delta  \rho}-\frac{\delta p_{r}}{\delta \rho} \sim v_{s \bot}^{2} - v_{s r}^{2}
\end{equation}\\
$v_{s \bot}^{2}$ and $ v_{s r}^{2}$ stand for the radial and tangential sound speeds, respectively \cite{abreu2007sound}.

This fundamental idea  employed in reviewing Herrera’s technique to identify possibly unstable anisotropic matter structures based on the concept of cracking. Now, by addressing the sound speeds and calculating Eq. (\ref{pertub}), we may not only have a more precise notion of the relative order of magnitude of the perturbations ($\delta \Delta$ and $\delta \rho$) but also what regions are more likely to be potentially unstable inside a matter configuration. There are two possible stable and two possible unstable regions: those where the gradient of anisotropy with respect to radial variable $r$ is larger than or equal to zero, and those where it is negative. It is clear that because $0\leq v_{sr}^{2} \leq 1$ and  $0\leq v_{s \bot}^{2} \leq 1$, we have $\mid v_{s \bot}^2-v_{s r}^2 \mid \leq 1$ \cite{abreu2007sound}. Hence,

\begin{align}
    -1 \leq v_{s\bot}^{2}-v_{s r}^2 \leq 1 \Rightarrow
        \begin{cases}
           -1 \leq v_{s\bot}^{2}-v_{s r}^2 \leq 0 & \text{potentially stable} \\
         0 < v_{s\bot}^{2}-v_{s r}^2 \leq 1 & \text{potentially unstable}
        \end{cases}
        \label{condition:1}
   \end{align}

\begin{figure}[h]
\centerline{\includegraphics[scale=0.4]{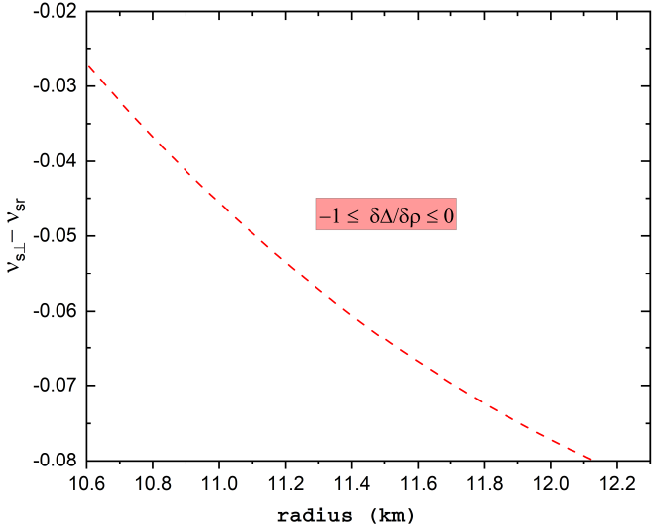}}
\caption{(Color online) Variation of the difference of radial and tangential sound speeds for the anisotropic envelope configuration. } 
\label{fig:pert}       
\end{figure}

Accordingly, we may now examine possible stable/unstable regions inside anisotropic models based on the difference of the propagation of sound within the matter configuration. Those regions where $v_{sr}^2 \geq v_{s \bot}^2$ will be  unstable. On the other hand, if $v_{s r}^{2}\leq v_{s \bot}^{2}$ everywhere within a matter distribution, no cracking will occur. For physically plausible models, the size of perturbations in the anisotropy should always be lower than those in density, i.e. $ \mid v_{s\bot}^{2} - v_{sr}^2 \mid $ $\leq $ $ 1 $ $\Rightarrow $ $\mid \delta \Delta \mid $ $\leq$  $\mid \delta \rho \mid$. When $\delta \Delta /\delta \rho > 0$, such perturbations lead to the possibility of unstable configuration. Profile for the $\delta \Delta / \delta \rho $ is displayed in Fig.\ref{fig:pert} for the anisotropic envelope region. For the TRV  core-envelope model under consideration , the perturbation relation $\delta \Delta / \delta \rho \equiv v_{s\bot}^{2}-v_{sr}^{2} $ satisfies the physical limit as given by  Eq.(\ref{condition:1}).

\begin{figure*}
    \centerline{
    \includegraphics[scale=0.24]{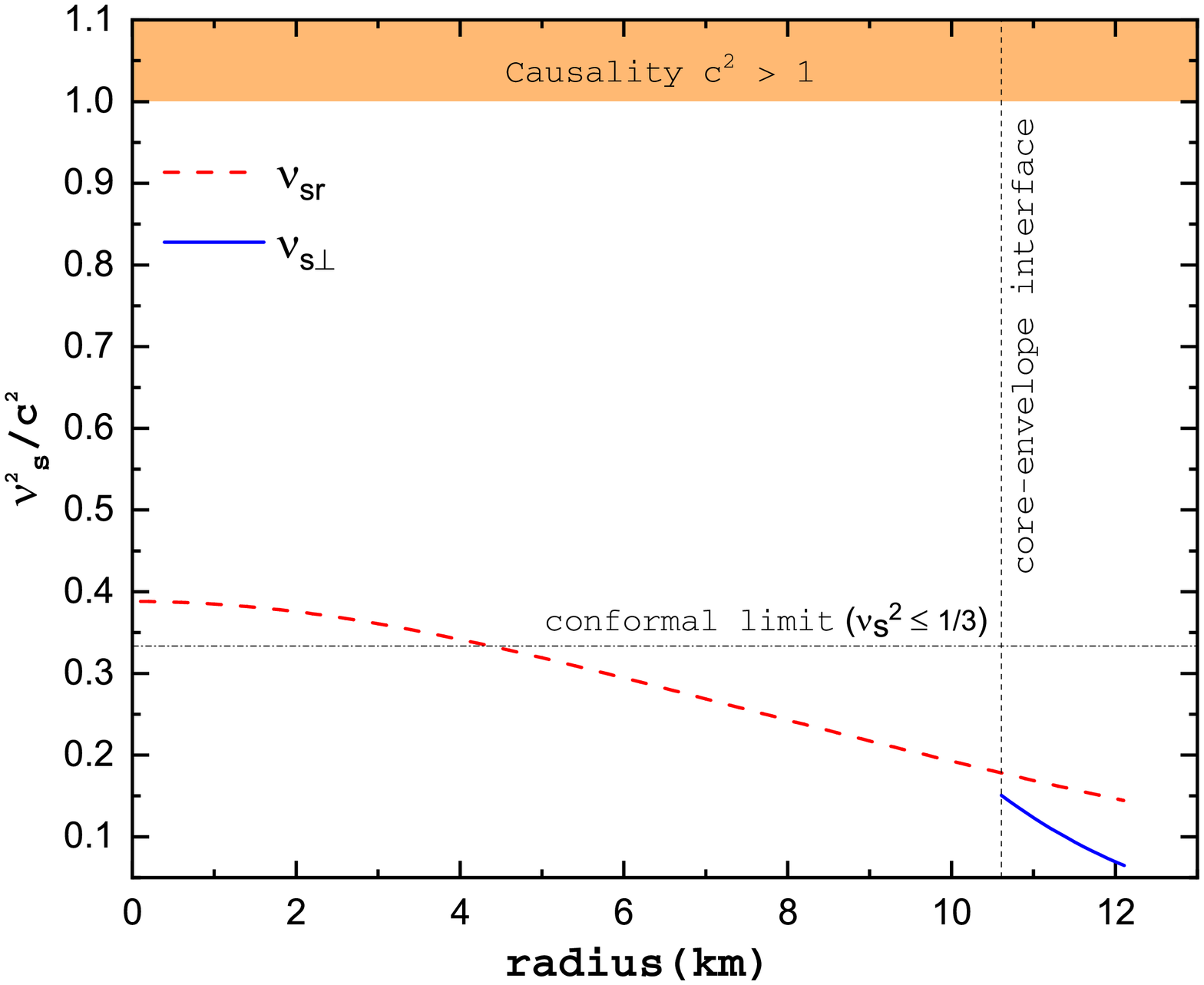}
    \includegraphics[scale=0.24]{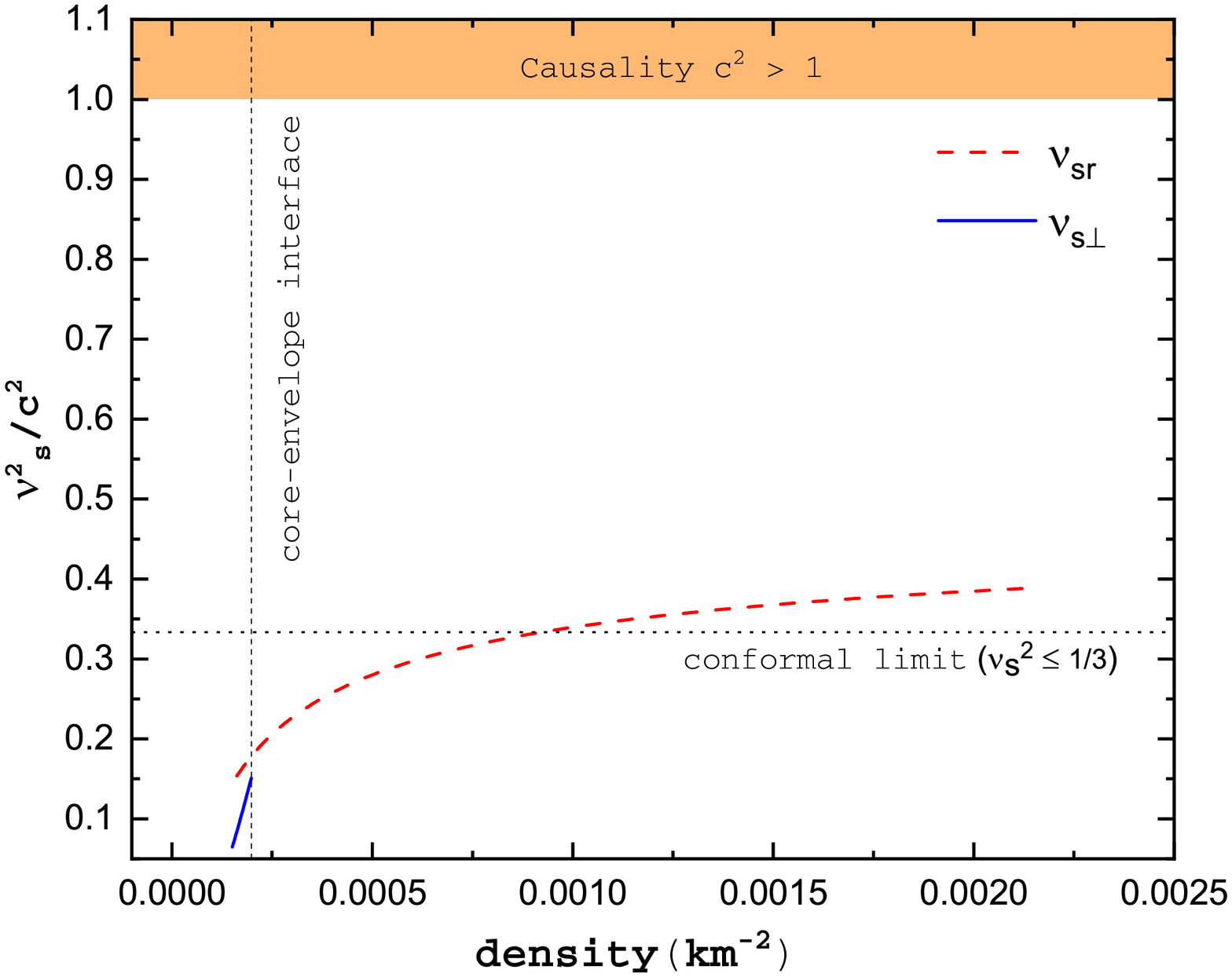}}
    \caption{(Color online) \textit{Left:} Speed of sound $v_{s}^2$ as a function radial distance for the TRV EoS. \textit{Right:} Speed of sound $v_{s}^2$ versus density. The vertical dotted line shows the core-envelope interface and the horizontal dotted line represents for the conformal limit ($v_{s}^{2} < 1/3$). The red solid  line represents radial velocity in the core and the envelope regions, respectively,. The solid blue line described tangential velocity in the envelope region, as well as grey upper band shows causality region.}
    \label{fig:vs}
\end{figure*}

In Fig. {\ref{fig:vs}}, the radial ($v_{sr}$) and tangential sound velocity ($v_{s \perp }$) as function of radii and density are shown.  From the computed results, we found that $v_{sr}$ as well as $v_{s \perp }$ is monotonically decreasing outward with the continuity at the interface for the $v_{sr}$ and discontinuity at interface for the $v_{s \perp }$. This discontinuity occurs at the core-envelope interface  actually comes from anisotropic pressure in the envelope region, and thus interferes with sound velocity. 

An important parameter which characterizes the stiffness of the EoS with respect to density perturbations is the \textit{adiabatic index} $\Gamma$.
The physical consistency of a relativistic anisotropic sphere is determined by  $\Gamma$, which is defined as the ratio of two specific heats \cite{bondi1964contraction}. The relativistic adiabatic indices  also are  important
parameters that affect the stability of any stellar
system. These are defined as

\begin{equation*}
    \Gamma_{r}(r) = \frac{p_{r}(r)+\rho(r)}{p(r)} \frac{\partial p_{r}(r)}{\partial \rho(r)}
\end{equation*}
\begin{equation}
    \Gamma_{\bot}(r) = \frac{p_{\bot}(r)+\rho(r)}{p_{\bot}(r)} \frac{\partial p_{\bot}(r)}{\partial \rho(r)}
\end{equation}

  

\begin{figure*}
    \centerline{
    \includegraphics[scale=0.24]{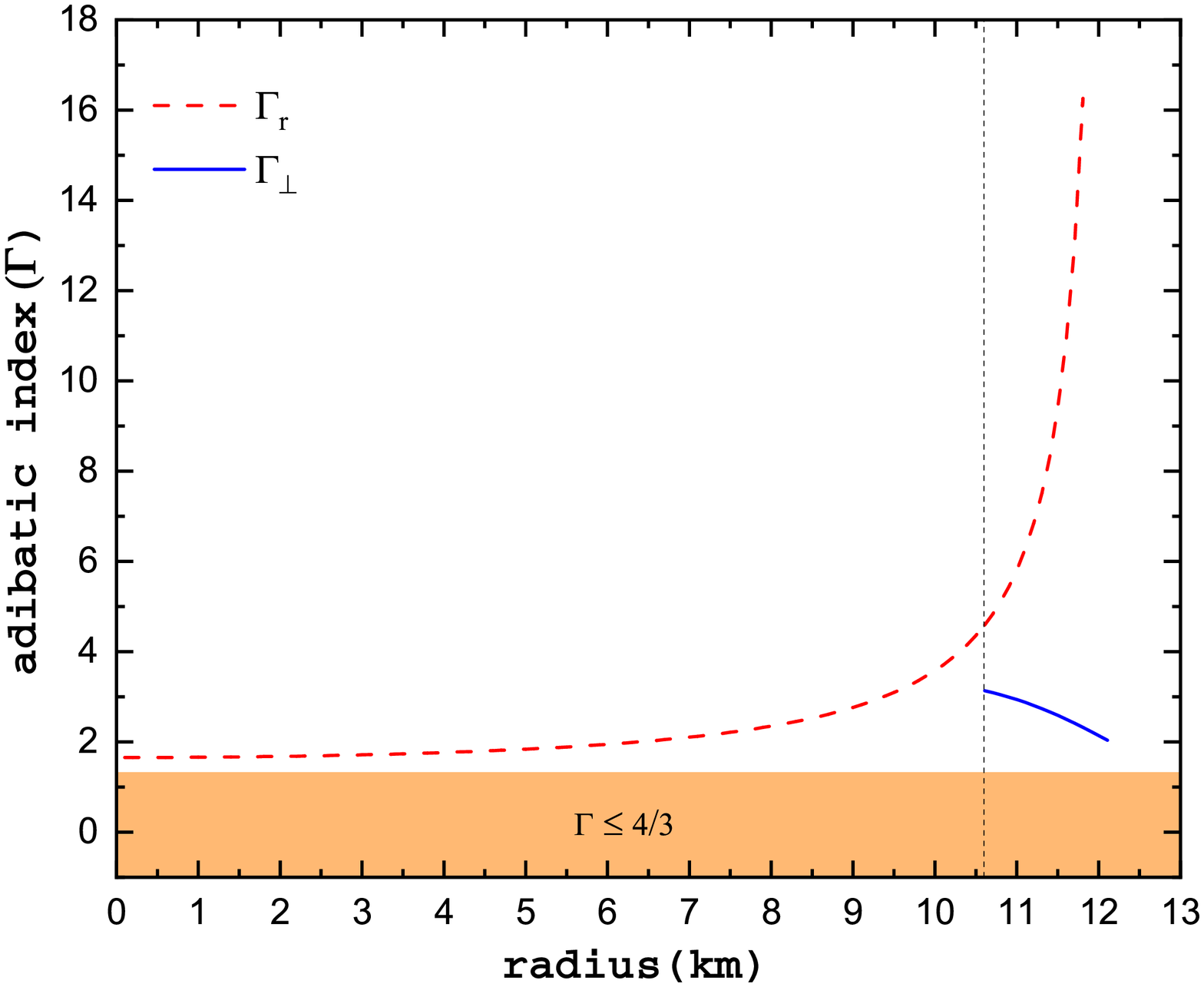}
    \includegraphics[scale=0.24]{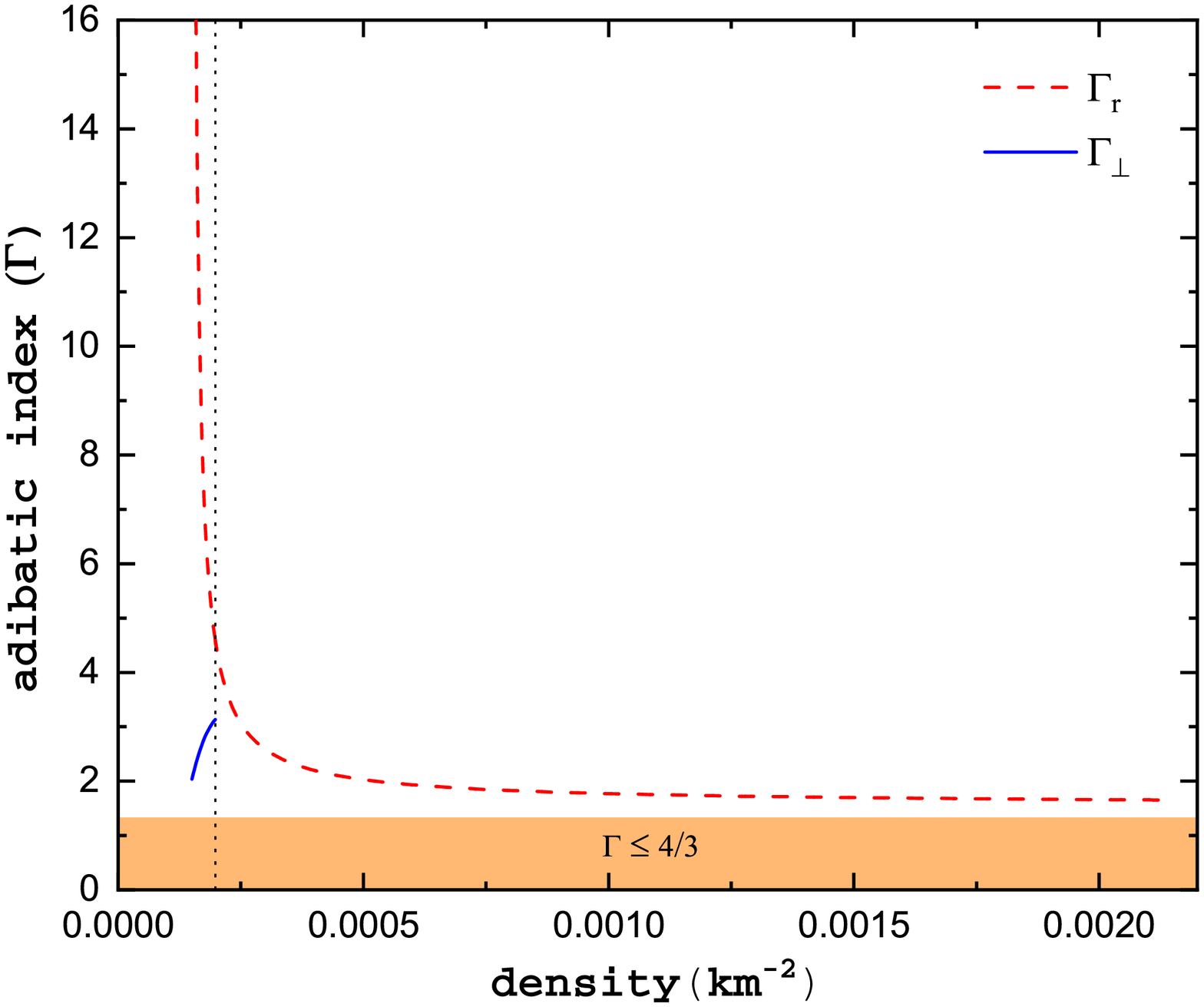}}
    \caption{(Color online) The adiabatic index $\Gamma$ versus radii and density(in geometrical unit) in a compact star core and envelope. The red-solid line is for radial $\Gamma$ in the core and evelope regions, respectively, the blue-solid line  corresponds to the case in which the tangential velocity the envelope region. The horizontal orange band is $\Gamma =4/3$ characteristic of a free ultra-relativistic Fermi gas. }
    \label{fig:gamma}
\end{figure*}

The profile of $\Gamma$ of the core and envelope of both pressures are plotted in Fig.{\ref{fig:gamma}}. Bondi \cite{bondi1964contraction} suggested that for a stable Newtonian sphere, the $\Gamma$ should be greater than $\frac{4}{3}$. The $\Gamma$ is a fundamental component of the instability criteria \cite{chandrasekhar1964dynamical}.  In particular, the amount that contains all of the main attributes of the equation of state on the instability formulas is described by Chandrasekhar \cite{chandrasekhar1964dynamical}, as a result, it serves as the link between the relativist structure of a spherical static object and the equation of state of the internal fluid. Specifically, $\Gamma$ varies from 2 to 4 in most equations of state of neutron star matter. More specifically, in certain circumstances,  $\Gamma$ is a weak function of density, in other cases, however, the density dependency is more complicated \cite{haensel2007neutron}.

The results of  $v_{s}^2$  and $\Gamma$ for the compact star matter are presented in  Fig.{\ref{fig:vs}} and Fig.{\ref{fig:gamma}}, respectively. The results show that there are possible stable regions inside the anisotropic envelope of the star. Because $\delta \Delta / \delta \rho < 0$, the sound speed stability criterion, Eq.(\ref{condition:1}) suggests that in the TRV anisotropic model, no cracking will occur.
Fig.{\ref{fig:gamma}} clearly indicate that the radial adiabatic indexes are continual  at the interface and satisfies the Bondi condition \cite{bondi1964contraction}, but in the tangential case it does not obey the Bondi condition. It will be the main reasons for the deformation and cracks for the Newtonian sphere.


\section{Estimates of stellar parameters $:$ $\Delta a$, $\Delta E_{g}$ and $\Delta I$}
\label{sec:5}
To account for the small anisotropy parameter, S of the order of $10^{-8}$ to $10^{-6}$, high numerical precision is required to arrive at the solution of the anisotropic TOV equation. Different anisotropic scales are presented here in the form of $\Delta$, and their values are chosen by the domain of anisotropy of Eq.(\ref{anisotropy}) (for $\lambda=0.07$).  In the case of isotropic pressure, $S=0$, Eq. (\ref{hydro}) turn out to be the TOV equation. We have incorporated $\Delta$ with Eq.(\ref{hydro}) for solving the anisotropic TOV equations,  we employ the TRV equation of state deduced from our earlier study of a core-envelope model of superdense stars \cite{khunt2021distinct,thomas2005core}. To begin with, the star's core is considered to be homogeneous, with the density, $\rho = \rho_{0}$. Here,  we have fixed the central density for the TRV EoS at $\rho _{0} = 1.34 \times 10^{15}$ g cm$^{-3}$, Eqs. (\ref{mass}) and  (\ref{hydro}) are integrated numerically to determine the global structure of (e.g. mass and radius) of a compact star.

The total gravitational energy ($E_{g}$) as well as the stellar moment of inertia ($I$) are computed as follows :

(i) \textit{Gravitational Energy}: 

A refined formula to compute the $E_{g}$ containing the \textit{ compactness parameter} $x_{GR}= r_{g}/a$, as proposed by \cite{lattimer2001neutron} is written as 

\begin{equation}
    E_{g} \simeq 1.6 \times 10^{53} \Big[\frac{M(\Delta)}{M_{\sun
}}\frac{x_{GR}(\Delta)}{0.3}\Big] \frac{1}{1-0.25 ~ x_{GR}(\Delta)} \,\, \text{erg}.
\end{equation}

Here, $x_{GR}  =2 G M/a c^{2}=r_{g}/a$, where $r_{g}$ is the Schwarzschild radius which depends implicitly on the anisotropic parameters, $\Delta$ through $M$ and $a$.
The importance of relativistic effects for a neutron star mass $M$ and radius $a$ is characterized by the  compactness parameter $r_{g}/a$.

(ii) \textit{Moment of inertia versus $M$ and $a$} :

 To quantify moment of inertia, we have used approximate formula relating $I$ to stellar mass and radius \cite{ravenhall1994neutron}. They showed that the ratio $I/M a^{2}$ depends mostly on the compactness parameter ($x_{GR}$) and is expressed as \cite{thorne1977relativistic}
 \begin{alignat}{2}
 I\simeq 0.21 M(\Delta) a^{2}_{\infty} ; \;\;\;\;
 a_{\infty}= \frac{a}{\sqrt{1-x_{GR}(\Delta)}} .
\end{alignat}
 where $a_{\infty}$ is an apparent radius related to the circumferential radius 

\begin{figure}[h]
\centerline{\includegraphics[scale=0.40]{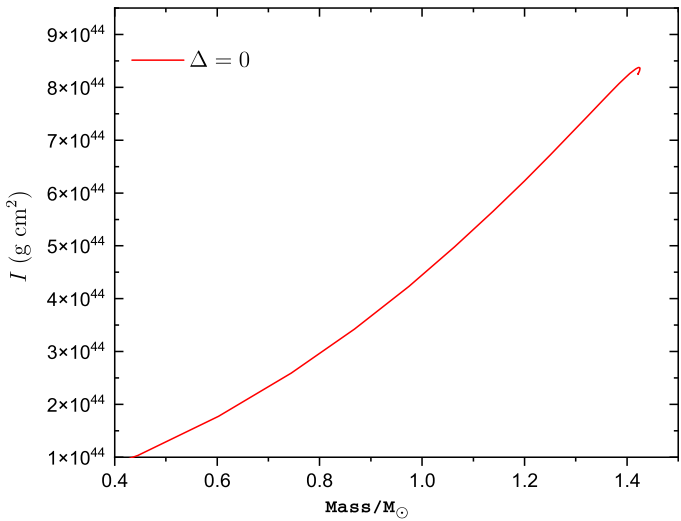}}
\caption{(Color online) Total moment of inertia versus total mass for $\lambda=0.07$ without anisotropy ($\Delta=0$).}
\label{fig:mi0}      
\end{figure}

In Fig.\ref{fig:mi0} we show the behaviour of the  moment of inertia with respect to the TOV mass for $\lambda=0.07$ with $\Delta=0$. The moment of inertia at maximum stable mass, $M_{max}$ is about $8.24 \times 10^{44}$ g cm$^2$.

In the computation, it is vital to enhance precision in order to accomplish the differences that we had in $ E_{g}(\Delta\neq 0)-E_{g}(\Delta=0)$ and $I(\Delta \neq 0)-I(\Delta=0)$. To optimize integration precision, we have divided the domain into smaller and smaller sections at the boundary, until the computed values are apparent.


\begin{figure}[h]
\centerline{ \includegraphics[scale=0.4]{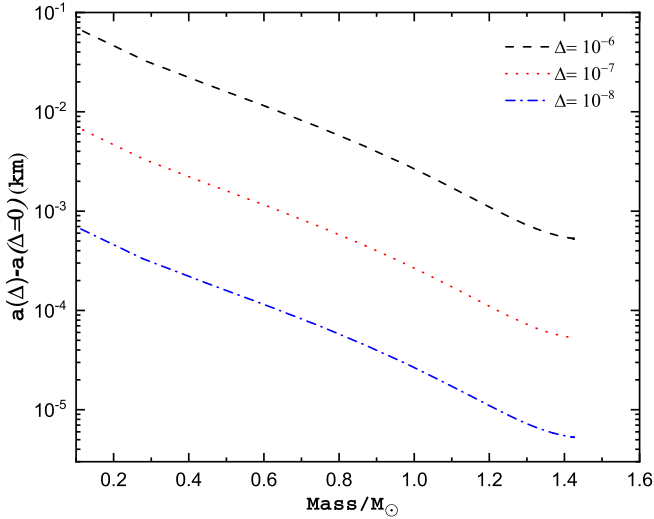}}
\caption{(Color online) The distinction $a (\Delta\neq 0)-a(\Delta = 0)$ in stellar radius as a function of stellar mass. Where black-dash line for $\Delta =10^{-6}$, red-dot line for $\Delta =10^{-7}$ and for $\Delta =10^{-8}$ is shown in blue dash-dot line.
\label{fig:2}}       
\end{figure}

\begin{figure}[h]
\centerline{ \includegraphics[scale=0.25]{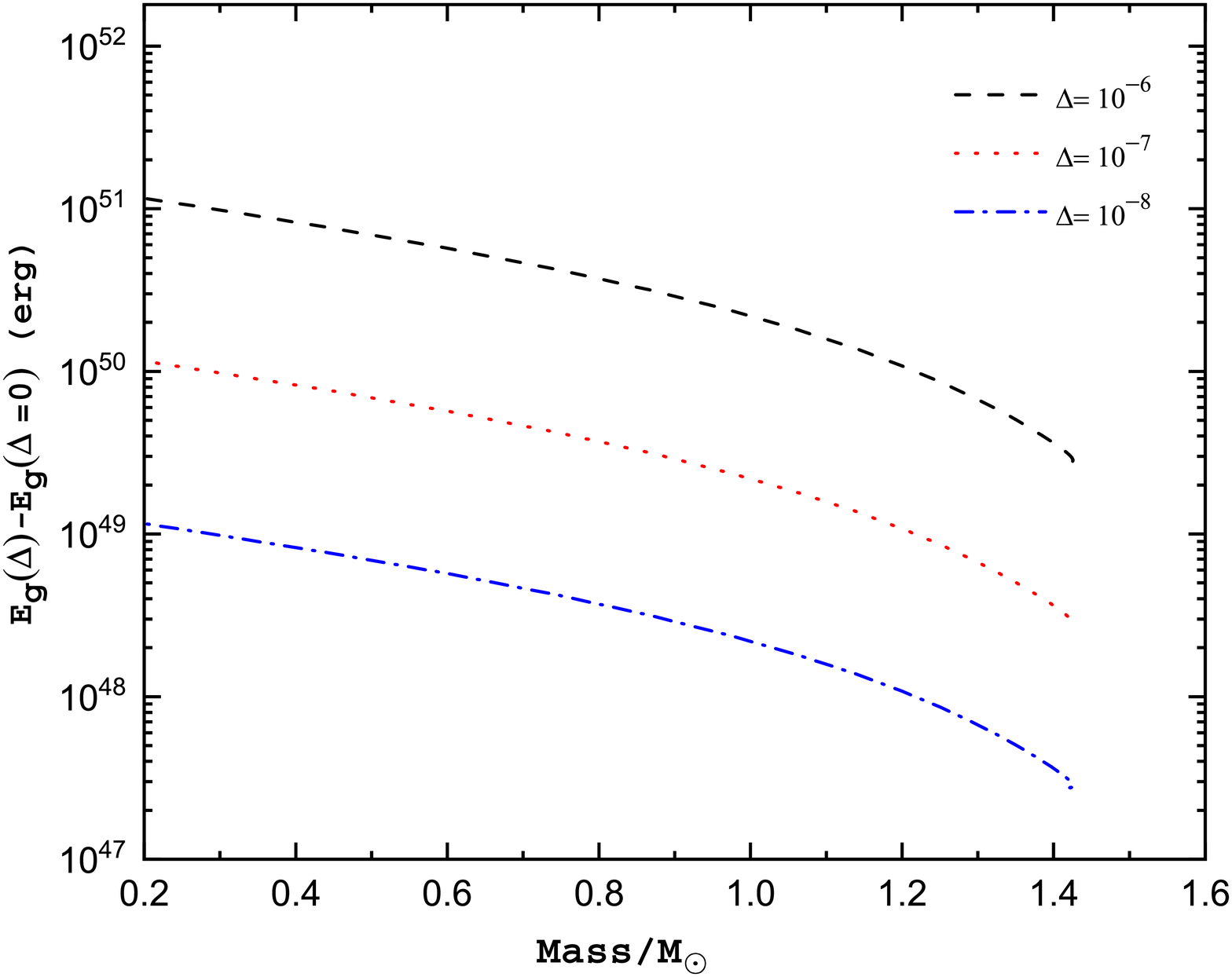}}
\caption{(Color online) As in Fig. \ref{fig:2}, but for difference in gravitational energy.
\label{fig:3}}      
\end{figure}

 \begin{figure}

\centerline{\includegraphics[scale=0.40]{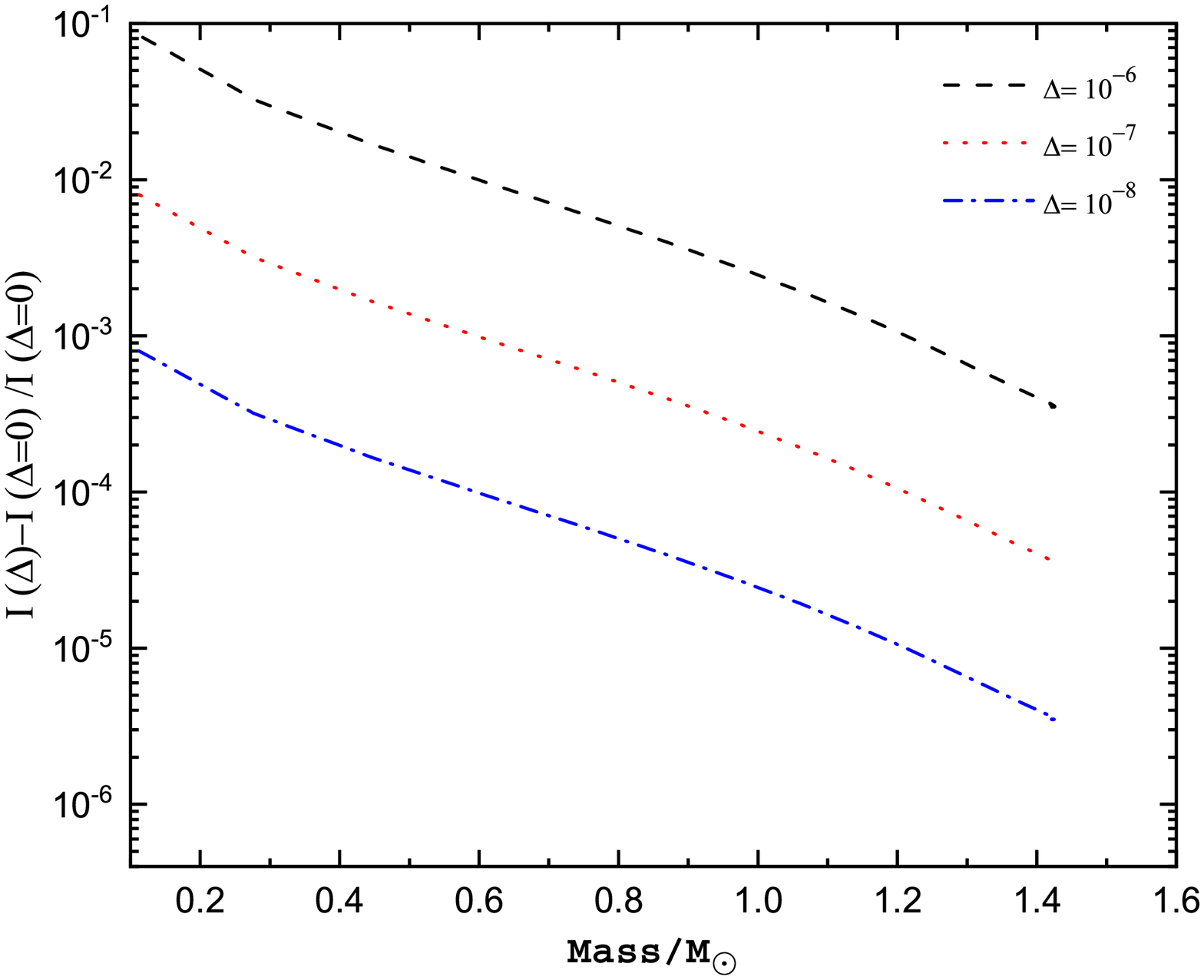}}
\caption{(Color online) As in Fig. \ref{fig:2}, but for the ratio difference in moment of inertia. 
\label{fig:4} }      
\end{figure}

The computed values of $a(M)$, $E_{g}(M)$ and $I(M)$ are shown in Fig. \ref{fig:2},~\ref{fig:3} and \ref{fig:4}, respectively. In the mass-radius calculation, we considered $\Delta$ from $10^{-6}$ to $10^{-8}$. From Fig.\ref{fig:2},  it is seen that as $\Delta$  varies from $10^{-8}$ to $10^{-6}$, the radius of the star in this range varies by $10^{-5}$ to $10^{-1}$ km.

Starquakes can cause a drastic change in the evelope region, releasing both gravitational and tangential strain energy \cite{franco2000quaking,horvath2005energetics,xu2003solid}. In general, the changes in stellar radius, $E_{g}$, and $I$ have direct correlation to stellar mass and anisotropic parameter $\Delta$. This indicates that in a highly compact self-bound  star with a greater mass, a quake-like event should be more significant for  larger change in $\Delta$.  Moreover, the disruptive instability of the radius may lead to more chaotic events at the core-envelope interface and  can cause huge  energy release. Fig.\ref{fig:3} shows the computed gravitational  energy difference corresponds to the anisotropy parameter $\Delta \sim 10^{-6} $ and $M_{max} \sim 1.43 M_{\sun}$. It is observed that the released energy is as high as $10^{50}$ erg. The energy released decreases with a decreasing magnitude of anisotropy (for $\Delta$ varies from $10^{-6}$ to $10^{-8}$). This scale of energy released  is comparable with the energy observed in  SGRs and GRBs (see in Table \ref{tab:1}). From our calculation as shown in Fig.\ref{fig:3}, we find that a giant starquake with $\Delta\leq 10^{-6}$ could produce such an energetic flare.  Based on different choices of the anisotropy parameters, \cite{xu2006superflares} have shown for quarks stars that the energy released on $\Delta= 10^{-5}$ is of the order of $10^{49}$ erg. A  change in $\Delta$ might cause a change in its spin frequency, $\Delta\Omega /  \Omega$ $= - \Delta I / I$ leading to glitches. We observed from Fig.\ref{fig:4} that such glitches with $\Delta \Omega \sim 10^{-5}$ to ${10^{-1}}$ could appear for mass variability of $M=0.2-1.43 M_{\sun})$ and $\Delta=10^{-8}$ to $10^{-6}$ and a giant energetic flare is considered to be associated with a high-amplitude glitch.

\section{ Discussion }
\label{sec:6}

To summarize, we have systematically investigated  the compact stars having a thin crust particularly its gravitational energy, the moment of inertia as an implicit function of the anisotropy magnitude ($\Delta$) predicted by the TRV core-envelope model of compact star with  anisotropy at the envelope region.
We modeled the compact star as a self-bound core of incompressible fluid with a fragile envelope.
The anisotropy variation as shown in  Fig.\ref{fig:1} indicates that the anisotropy starts at the core boundary  increases up to a radial distance of 10 to 12 km . It  further decreases towards  the envelope boundary of the star. In Fig.\ref{fig:2}, we have shown the mass-radius relation for different values of the anisotropy parameter in the range $10^{-8} \leq \Delta \leq 10^{-6}$  and computed the difference in the radius of the star.  Further the gravitational energy differences and moment of inertia difference due to the  radial fluctuations caused by the anisotropy parameter  are also estimated (see Fig.\ref{fig:3}). It is found that the magnitude of this energy is comparable to the energy produced by SGRs and GRBs.

Cracking is a concept that was first suggested for self-gravitating anisotropic matter configurations by L Herrera and collaborators \cite{herrera1992cracking,di1994tidal,di1997cracking}. This concept has been reintroduced in our analysis. In this study, we followed their approach and studied the cracking and stability of self-bound anisotropic stars. It has been revealed that, for certain dependent perturbation in particular, the ratio for fluctuations in anistropy to energy density, $\delta \Delta / \Delta \rho$, could be well understood in terms of the difference between the velocities of sound (i.e. $\delta \Delta / \delta \rho \equiv v_{s\bot}^{2}-v_{sr}^{2} $). Cracking points in a configuration that satisfies $\mid \delta \Delta/ \delta \rho \mid > 1$, would not lead to physically unstable models. From the present study, we are able to clearly determine, based on equation (\ref{condition:1}), the region within the matter distribution are more likely to be potentially stable/unstable. From the Fig.\ref{fig:pert}, we see that $ -1 \leq v_{s\bot}^{2}-v_{s r}^2 \leq 0 $
 condition is satisfied in the anisotropic configuration. Our results indicate that anisotropic structure is potentially stable (as shown in Fig.\ref{fig:pert}). Moreover the work done by  \cite{abreu2007sound}  argued that when the system is entirely anisotropic ($p_{\bot}\neq 0$ and $p_{r} =0$) the system may be considered possibly stable and it can be essential for cracking like event. The faults generated due to the cracking events in the evelope region can lead to quakes. The superflares of $\gamma$-ray emission (e.g., GRBs, SGRs) could be caused by  quakes in such cases.

\section{Conclusion}\label{conclusion}
In conclusion, we are able to propose a viable model towards our understanding of the production of GRBs to the conventional framework. According to our theoretical calculation and numerical results : if the tangential pressure is slightly larger [only $(1+10^{-6})$ times] than the radial pressure in a self bound star with mass $1.43 M_{\sun}$, the released gravitational energy during a quake could be as high as $10^{50}$ erg, for spherically symmetric compact stars. We have examined the evolution of anisotropy in the envelope of highly compact star. 
 The envelope develops cracks under the pressure stress caused by  density fluctuations induced by the anisotropy. The amount of energy  emitted from this event is very remarkable.
It is possible for stresses to build in the envelope of a high compact star resulting  the crust being fractured (i.e., a starquake), which may have an impact on the star's spin development and generate  glitches.

A star-quake is a possible activation source for stellar instabilities, e.g., it might be linked to a pulsar glitch or a magnetar flare. The idea is that the star's evolution, such as magnetic field loss, produces strain in the elastic crust, and  the stored energy is abruptly released when the system reaches a critical threshold. The usual energy released in this process would consequently be  of the order of the maximum mechanical energy that can be stored in the crust, which is estimated to be $10^{-9}-10^{-7} M_{\sun} c^{2}$ \cite{blaes1989neutron,mock1998limits}. This  is significant considering magnetars appear to be associated with rather regular flare outbursts \cite{watts2016colloquium}. Internal phase transitions are another possibility linked with the star's evolution.
From the point of view of the present study , it is important to note that the anisotropy largely at the envelope region that causes deformation at the envelope (star crust) leading to matter being thrown out and falling back.

\section*{Declarations}
\begin{itemize}
    \item Funding : Not applicable
    \item Conflict of interest/Competing interests: Not applicable
    \item Availability of data and materials: No data has been used in this paper and the materials used have been cited in appropriate places
    \item  Code availability: Not applicable
    \item  Authors’  contributions:  (i)  Khunt:  Conceptualization,  methodology, Analytical and Numerical calculation, manuscript preparation ; (ii) Thomas :  investigation, writing-review, ; (iii) Vinodkumar:  writing-review, calculation-analysis
\end{itemize}

\section{Appendix A : Anisotropy profile} \label{app_1}
In this appendix, we show anisotropy plot for values of $\lambda=$ 0.1, 0.3, 0.5, and 0.9 from the TRV model presented in Subsection \ref{envelope_sec}.

\begin{figure}[H]
\centerline{ \includegraphics[scale=0.4]{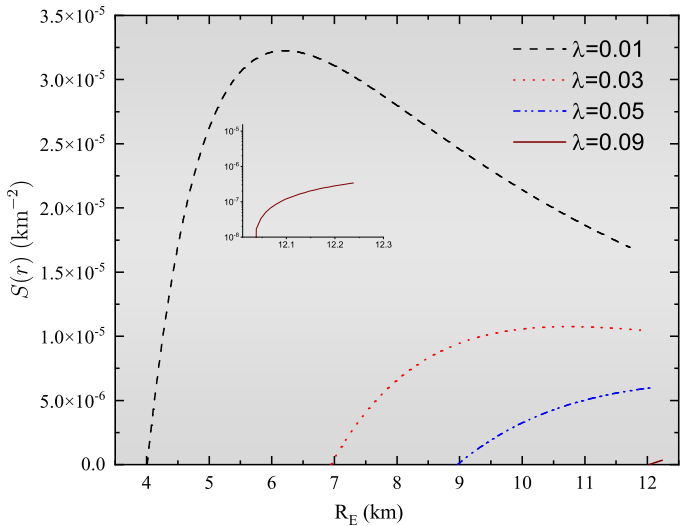}}
\caption{(Color online) Variation of an anistropy $S$ in km$^{-2}$ with respect to a envelope radius of the star. For a density variation of  $\;\; \lambda =0.1, 0.3, 0.5 \;\; \text{and}, 0.9$.
\label{fig:s_all}}       
\end{figure}

\bibliographystyle{ijpbst}
\bibliography{sn-bibliography}

\end{document}